\documentclass[preprint,12pt]{elsarticle} 
\usepackage{graphicx} 
\usepackage{amssymb} 
\usepackage{amsmath} 
\usepackage{amsfonts}  
\usepackage{slashed} 
\usepackage[dvipsnames]{xcolor} 
\def\lamb#1#2{$^{#1}_{\Lambda}${#2}}
\def\lam#1#2{$^{#1}_{~\Lambda}${#2}}

\newcommand{\be}{\begin{equation}} 
\newcommand{\ee}{\end{equation}}

\journal{Physics Letters B}

\begin{document} 

\begin{frontmatter}

\title{Constraints on $\Xi^-$ nuclear interactions from capture events 
in emulsion} 

\author{~E.~Friedman}
\author{~A.~Gal\corref{cor1}~}
\address{Racah Institute of Physics, The Hebrew University, Jerusalem
91904, Israel}
\cortext[cor1]{corresponding author: Avraham Gal, avragal@savion.huji.ac.il}

\date{\today} 

\begin{abstract} 
 
Five $\Xi^- p\to \Lambda\Lambda$ two-body capture events in $^{12}$C and 
$^{14}$N emulsion nuclei, in which a pair of single-$\Lambda$ hypernuclei is 
formed and identified by their weak decay, have been observed in $(K^-,K^+)$ 
emulsion exposures at KEK and J-PARC. Applying a $\Xi^-$-nucleus optical 
potential methodology to study atomic and nuclear transitions, we confirm 
that these capture events occur from Coulomb assisted $1p_{\Xi^-}$ nuclear 
states. Long-range $\Xi N$ shell-model correlations are found essential to 
achieve consistency between the $^{12}$C and $^{14}$N events. The resulting 
$\Xi$-nuclear interaction is strongly attractive, with $\Xi$ potential depth 
in nuclear matter $V_{\Xi}\gtrsim 20$~MeV. Implications to multi-strangeness 
features of dense matter are outlined. 

\end{abstract}


\end{frontmatter}

\section{Introduction and background} 
\label{sec:intro} 

Recent two-particle correlation studies of $p\Lambda$, $\Lambda\Lambda$ and 
$\Xi^-p$ pairs measured by ALICE~\cite{ALICE19a,ALICE19b,ALICE19c,ALICE20} in 
$pp$ and $p$-Pb ultra-relativistic collisions at TeV energies have triggered 
renewed interest in Strangeness ${\cal S}\neq 0$ baryon-baryon interactions 
and consequences thereof to strange hadronic matter. In particular, 
the $\Xi^-p$ interaction was shown to be attractive~\cite{ALICE19c}, 
in good agreement with the recent HAL-QCD lattice calculations reaching 
$m_{\pi}=146$~MeV~\cite{HALQCD20}. Understanding the strength of the 
${\cal S}=-2$ $\Xi N$ interaction, particularly when embedded in nuclear 
media, is vital for resolving the Hyperon Puzzle which addresses the fate 
of hyperons in dense neutron-star matter~\cite{TF20}. 

Little is known from experiment on the nuclear interaction of $\Xi$ 
hyperons~\cite{GHM16,HN18}. A standard reaction production is the 
nuclear $(K^-,K^+)$ reaction, driven by $K^-p\to K^ + \Xi^-$ strangeness 
exchange on protons. Owing to its large momentum transfer, the produced 
$\Xi^-$ hyperons populate dominantly the quasi-free continuum region, with 
less than 1\% expected to populate $\Xi^-$-nuclear bound states that decay 
subsequently by the $\Xi^- p\to\Lambda\Lambda$ strong-interaction capture 
reaction. Analysis of old emulsion events attributed to the 
formation of $\Xi$ hypernuclei suggested attractive Woods-Saxon $\Xi$ nuclear 
potential depth $V_{\Xi}$=21-24~MeV~\cite{DG83}. While this range of values 
is considered sufficient for $\Xi$ hyperons to play an active role in strange 
hadronic matter~\cite{Schaffner00} and in neutron stars~\cite{Schaffner12}, 
somewhat smaller values follow from studies of dedicated $(K^-,K^+)$ 
counter experiments: $V_{\Xi}\lesssim 20$~MeV in KEK PS-E224~\cite{E224}, 
$V_{\Xi}\sim 14$~MeV in BNL AGS-E885~\cite{E885}, both on $^{12}$C, and 
$V_{\Xi}=17\pm 6$~MeV on $^9$Be~\cite{HH21} from BNL AGS-E906~\cite{E906}. 
New results from the J-PARC E05 and E70 $(K^-,K^+)$ experiments on $^{12}$C 
are forthcoming~\cite{E05}. However, no $\Xi^-$ or $\Lambda\Lambda$ 
hypernuclear bound states have ever been observed unamiguously in these 
experiments. Powerful future experiments by the PANDA Collaboration at 
FAIR~\cite{pocho17}, using ${\bar p}p\to \Xi^-{\bar \Xi}^+$ or ${\bar p}n\to 
\Xi^-{\bar \Xi}^0$ production modes, and also at BESIII~\cite{YK21} focusing 
on the $J/\psi \to\Xi^- {\bar \Xi}^+$ $O(10^{-3})$ decay branch, are likely 
to change this state of the art. 

The situation is different in exposures of light-emulsion CNO nuclei to the 
$(K^-,K^+)$ reaction, in which a tiny fraction of the produced high-energy 
$\Xi^-$ hyperons slow down in the emulsion, undergoing an Auger process to 
form high-$n$ atomic levels, and cascade down radiatively. Cascade essentially 
terminates, when strong-interaction capture takes over, in a 3D atomic state 
bound by 126, 175, 231 keV in C, N, O, respectively. The 3D strong-interaction 
shift is less than 1~keV~\cite{BFG99}. Capture events are recorded by 
observing $\Lambda$ hyperon or hypernuclear decay products. Interestingly, 
whereas the few observed double-$\Lambda$ hypernucleus production events are 
consistent with $\Xi^-$ capture from atomic 3D states~\cite{HN18}, formation 
of pairs of single-$\Lambda$ hypernuclei requires capture from a lower $\Xi^-$ 
orbit. Expecting the final two $\Lambda$ hyperons in $\Xi^-p\to \Lambda\Lambda
$ capture to be formed in a $S=0,\,1s_{\Lambda}^2$ configuration, the initial 
$\Xi^-$ hyperon and the proton on which it is captured must satisfy $l_{\Xi^-}
=l_p$~\cite{Zhu91} which for $p$-shell nuclear targets favors the choice 
$l_{\Xi^-}=1$. Indeed, all two-body $\Xi^-$ capture events, $\Xi^-+{^{A}Z}\to
$~\lam{A'}{Z'}+\lam{A''}{Z''}, to twin single-$\Lambda$ hypernuclei reported 
from KEK and J-PARC light-emulsion $K^-$ exposures~\cite{E176,E373,E07a}, 
as listed here in Table~\ref{tab:events}, are consistent with $\Xi^-$ capture 
from Coulomb-assisted $1p_{\Xi^-}$ nuclear states bound by $\sim$1~MeV. 

\begin{table}[htb] 
\begin{center} 
\caption{Reported two-body $\Xi^-$ capture events 
$\Xi^-+{^{A}Z}\to$~\lam{A'}{Z'}+\lam{A''}{Z''} in light-emulsion nuclei to a 
pair of single-$\Lambda$ hypernuclei, some in ground states, some in specific 
excited states marked by asterisk. Only the first and last events are uniquely 
assigned. Fitted $\Xi^-$ binding energies $B_{\Xi^-}$ are listed.}  
\begin{tabular}{ccccc} 
\hline   
Experiment & Event & $^{A}Z$ & \lamb{A'}{Z'}+\lamb{A''}{Z''} & $B_{\Xi^-}$ 
(MeV)  \\ 
\hline 
KEK E176 \cite{E176} & 10-09-06 & $^{12}$C & \lamb{4}{H}+\lamb{9}{Be} & 
0.82$\pm$0.17  \\ 
KEK E176 \cite{E176} & 13-11-14 & $^{12}$C & \lamb{4}{H}+\lamb{9}{Be}$^{\ast}$ 
& 0.82$\pm$0.14  \\
KEK E176 \cite{E176} & 14-03-35 & $^{14}$N & \lamb{3}{H}+\lam{12}{B} & 
1.18$\pm$0.22  \\ 
KEK E373 \cite{E373} & KISO & $^{14}$N & \lamb{5}{He}+\lam{10}{Be}$^{\ast}$ & 
1.03$\pm$0.18  \\ 
J-PARC E07 \cite{E07a} & IBUKI & $^{14}$N & \lamb{5}{He}+\lam{10}{Be} & 
1.27$\pm$0.21  \\  
\hline 
\end{tabular}  
\label{tab:events} 
\end{center} 
\end{table} 

In Table~\ref{tab:events} only the first and last listed events are uniquely 
assigned in terms of initial emulsion nucleus $^{A}Z$ and final single-$
\Lambda$ hypernuclei \lamb{A'}{Z'}+\lamb{A''}{Z''} ground states, providing 
thereby a unique value of $B_{\Xi^-}$ per each event. The events fitted by 
assuming specific excited states \lamb{A''}{Z''}$^{\ast}$ are equally well 
fitted each by g.s. assignments \lamb{A''}{Z''}, with values of $B_{\Xi^-}$ 
as high as $\sim$4~MeV, and the middle event is equally well fitted as 
a capture event in $^{16}$O, to \lamb{3}{H}+\lam{14}{C} with $B_{\Xi^-}=0.46
\pm 0.39$~MeV or to \lamb{4}{H}+\lam{13}{C} with $B_{\Xi^-}=0.40\pm0.27$~MeV, 
both consistent with $\Xi^-$ capture from an atomic 3D state. We note that 
the listed $\Xi^-$ binding energy $B_{\Xi^-}$ values are all around 1~MeV, 
significantly higher than the purely-Coulomb atomic 2P binding energy values 
which are approximately 0.3, 0.4, 0.5 MeV in C,N,O atoms, respectively. 
These $\sim$1~MeV $B_{\Xi^-}$ values correspond to $1p_{\Xi^-}$ nuclear 
states that evolve from 2P atomic states upon adding a strong-interaction 
$\Xi$ nuclear potential.{\footnote{Nuclear single-particle (s.p.) states 
are denoted by lower-case letters: $1s,1p,1d,...$ for the lowest $l$ values, 
in distinction from atomic s.p. states denoted by capitals: 1S,2P,3D,... 
for the lowest L values.}} This interpretation is the only one common to 
{\it all} five events. 

Not listed in the table are multi-body capture events that require for 
their interpretation undetected capture products, usually neutrons, on top 
of a pair of single-$\Lambda$ hypernuclei. Most of these new J-PARC E07 
events~\cite{E07b} imply $\Xi^-$ capture from $1s_{\Xi^-}$ nuclear states, 
with capture rates ${\cal O}(10^{-2})$ of capture rates from the $1p_{\Xi^-}$ 
nuclear states considered here~\cite{Zhu91,Koike17}. 

In the present work we consider $\Xi^-$ atomic and nuclear transitions in 
light emulsion atoms, first with a $t\rho$ optical potential, to substantiate 
that $\Xi^- p\to\Lambda\Lambda$ capture indeed occurs from a Coulomb-assisted 
nuclear $1p_{\Xi^-}$ state in light emulsion nuclei. The strength of this 
$\Xi$-nuclear potential is determined by requiring that it reproduces 
a $1p_{\Xi^-}$ state in $^{12}$C(0$^+_{\rm g.s.}$) bound by 0.82$\pm$0.15~MeV 
from Table~\ref{tab:events}. Disregarding temporarily the $s_{\Xi^-}=
\frac{1}{2}$ Pauli-spin degree of freedom, we proceed to discuss the 
implications of identifying the value $B_{\Xi^-}\approx 1.15\pm 0.20$~MeV 
for $^{14}$N from Table~\ref{tab:events} with the binding energy of 
${\cal L}^{\pi}=(0^-,1^-,2^-)$ triplet of $1p_{\Xi^-}$ nuclear states 
built on $J^{\pi}(^{14}$N$_{\rm g.s.}$)=1$^+$, thereby linking the capture 
process to properties of the $\Xi$-nucleus residual interaction. This provides 
the only self consistent deduction of the $\Xi$-nuclear interaction strength 
from analysis of the five $\Xi^-$ capture events in light nuclear emulsion, 
fitted to two-body formation of specific $\Lambda$ hypernuclei, as listed 
in Table~\ref{tab:events}. The resulting $t\rho$ $\Xi$ potential depth at 
nuclear-matter density $\rho_0$=0.17~fm$^{-3}$ is $V_{\Xi}\approx 24$~MeV. 
We then improve upon the $t\rho$ leading term of the optical potential by 
introducing the next, Pauli correlation term in the optical potential density 
expansion~\cite{DHL71}. This leads to $\approx$10\% reduction of $V_{\Xi}$, 
down to $V_{\Xi}\approx 22$~MeV, keeping it within the optical potential 
approach well above 20~MeV. Our results suggest that $1s_{\Xi^-}$ nuclear 
bound states exist all the way down to $^4$He, with potentially far-reaching 
implications for the role of $\Xi$ hyperons in multi-strange dense matter. 

A value $V_{\Xi}\gtrsim 20$~MeV implies a substantially stronger in-medium 
$\Xi N$ attraction than reported by some recent model evaluations 
(HAL-QCD~\cite{HALQCD19}, EFT@NLO~\cite{HM19,K19}, RMF~\cite{Gaitanos21}) 
all of which satisfy $V_{\Xi}\lesssim 10$~MeV. A notable exception is provided 
by versions ESC16*(A,B) of the latest Nijmegen extended-soft-core $\Xi N$ 
interaction model~\cite{ESC16}, in which values of $V_{\Xi}$ higher than 
20~MeV are derived. However, these large values are reduced substantially 
by $\Xi NN$ three-body contributions within the same ESC16* model.

\section{Methodology} 
\label{sec:meth} 

The starting point in optical-potential analyses of hadronic 
atoms~\cite{FG07} is the in-medium hadron self-energy $\Pi(E,\vec p,\rho)$ 
that enters the in-medium hadron (here $\Xi$ hyperon) dispersion relation
\begin{equation} 
E^2-{\vec p}^{~2}-m_{\Xi}^2-\Pi(E,\vec p,\rho)=0, 
\label{eq:disp} 
\end{equation}
where ${\vec p}$ and $E$ are the $\Xi$ momentum and energy, respectively,
in nuclear matter of density $\rho$. The resulting $\Xi$-nuclear optical
potential $V_{\rm opt}$, defined by $\Pi(E,\vec p,\rho)=2EV_{\rm opt}$,
enters the near-threshold $\Xi^-$ wave equation
\begin{equation} 
\left[ \nabla^2  - 2{\mu}(B+V_{\rm opt} + V_c) + (V_c+B)^2\right] \psi = 0, 
\label{eq:KG} 
\end{equation}
where $\hbar = c = 1$. Here $\mu$ is the $\Xi^-$-nucleus reduced mass, 
$B$ is the complex binding energy, $V_c$ is the finite-size Coulomb potential 
of the $\Xi^-$ hyperon with the nucleus, including vacuum-polarization terms, 
all added according to the minimal substitution principle $E \to E - V_c$. 
Strong-interaction optical-potential $V_{\rm opt}$ terms other than 
$2\mu V_{\rm opt}$ are negligible and omitted here. The use of a Klein-Gordon 
wave equation (\ref{eq:KG}) for the $\Xi^-$ fermion rather than Dirac equation 
provides an excellent approximation when $Z\alpha\ll 1$ and fine-structure 
effects are averaged on, as for the X-ray transitions considered here. 
$\Xi^-$ nuclear spin-orbit effects are briefly mentioned below. 

For $V_{\rm opt}$ in Eq.~(\ref{eq:KG}) we first use a standard $t\rho$ 
form~\cite{FG07} 
\begin{equation} 
V_{\rm opt}(r)=-\frac{2\pi}{\mu}(1+\frac{A-1}{A}\frac{\mu}{m_N})
[b_0\rho(r)+b_1\rho_{\rm exc}(r)], 
\label{eq:EEs} 
\end{equation} 
where the complex strength parameters $b_0$ and $b_1$ are effective, 
generally density dependent $\Xi N$ isoscalar and isovector c.m. scattering 
amplitudes respectively. The density $\rho=\rho_n+\rho_p$ is a nuclear 
density distribution normalized to the number of nucleons $A$ and 
$\rho_{\rm exc}=\rho_n-\rho_p$ is a neutron-excess density with 
$\rho_n=(N/Z)\rho_p$, implying that $\rho_{\rm exc}=0$ for the $N=Z$ 
emulsion nuclei $^{12}$C and $^{14}$N analyzed in the next section. 
Here we used mostly nuclear density distributions of harmonic-oscillator 
(HO) type~\cite{Elton61} where the r.m.s. radius of $\rho_p$ was set equal 
to that of the known nuclear charge density~\cite{AM13}. Folding reasonably 
chosen $\Xi N$ interaction ranges other than corresponding to the proton 
charge radius, or using Modified Harmonic Oscillator (MHO) densities, 
or replacing HO densities by realistic three-parameter Fermi (3pF) density 
distributions~\cite{JVV74,VJV87} made little difference: all the calculated 
binding energies changed by a small fraction, about 0.03~MeV of the 
uncertainty imposed by the $\pm$0.15 MeV experimental uncertainty 
of the 0.82~MeV $1p_{\Xi}$ binding energy in $^{12}$C listed in 
Table~\ref{tab:events}. We note that the central density $\rho(0)$ in all 
density versions used here is within acceptable values for nuclear matter, 
i.e., between roughly 0.15 and 0.20 fm$^{-3}$. 

The form of $V_{\rm opt}$ given by Eq.~(\ref{eq:EEs}) corresponds to a 
central-field approximation of the $\Xi$-nuclear interaction. Spin and 
isospin degrees of freedom induced by the most general two-body $s$-wave 
$\Xi N$ interaction $V_{\Xi N}$, 
\begin{equation} 
V_{\Xi N}=V_0+V_{\sigma}\sigma_{\Xi}\cdot\sigma_N+V_{\tau}\tau_{\Xi}\cdot
\tau_N+V_{\sigma\tau}\sigma_{\Xi}\cdot\sigma_N\,\tau_{\Xi}\cdot\tau_N 
\label{eq:V2body} 
\end{equation} 
with $V$s functions of $r_{\Xi N}$, are suppressed in this approach. Choosing 
$^{12}$C$_{\rm g.s.}$ with isospin $I=0$ and spin-parity $J^{\pi}=0^+$ for a 
nuclear medium offers the advantage that only $V_0$ is operative in leading 
order since the nuclear expectation value of each of the other three terms in 
Eq.~(\ref{eq:V2body}) vanishes. But for $^{14}$N$_{\rm g.s.}$, with 
$I(J^{\pi})$=0(1$^+$), $V_{\sigma}$ is operative as well, adding unavoidable 
model dependence of order ${\cal O}(1/A)$ to $\Xi^-$-nuclear potential depth 
values derived from capture events assigned to this emulsion nucleus. For this 
reason we start the present analysis with the two $\Xi^-$-$^{12}$C emulsion 
events of Table~\ref{tab:events}. 

\begin{figure}[htb] 
\begin{center}  
\includegraphics[width=0.8\textwidth]{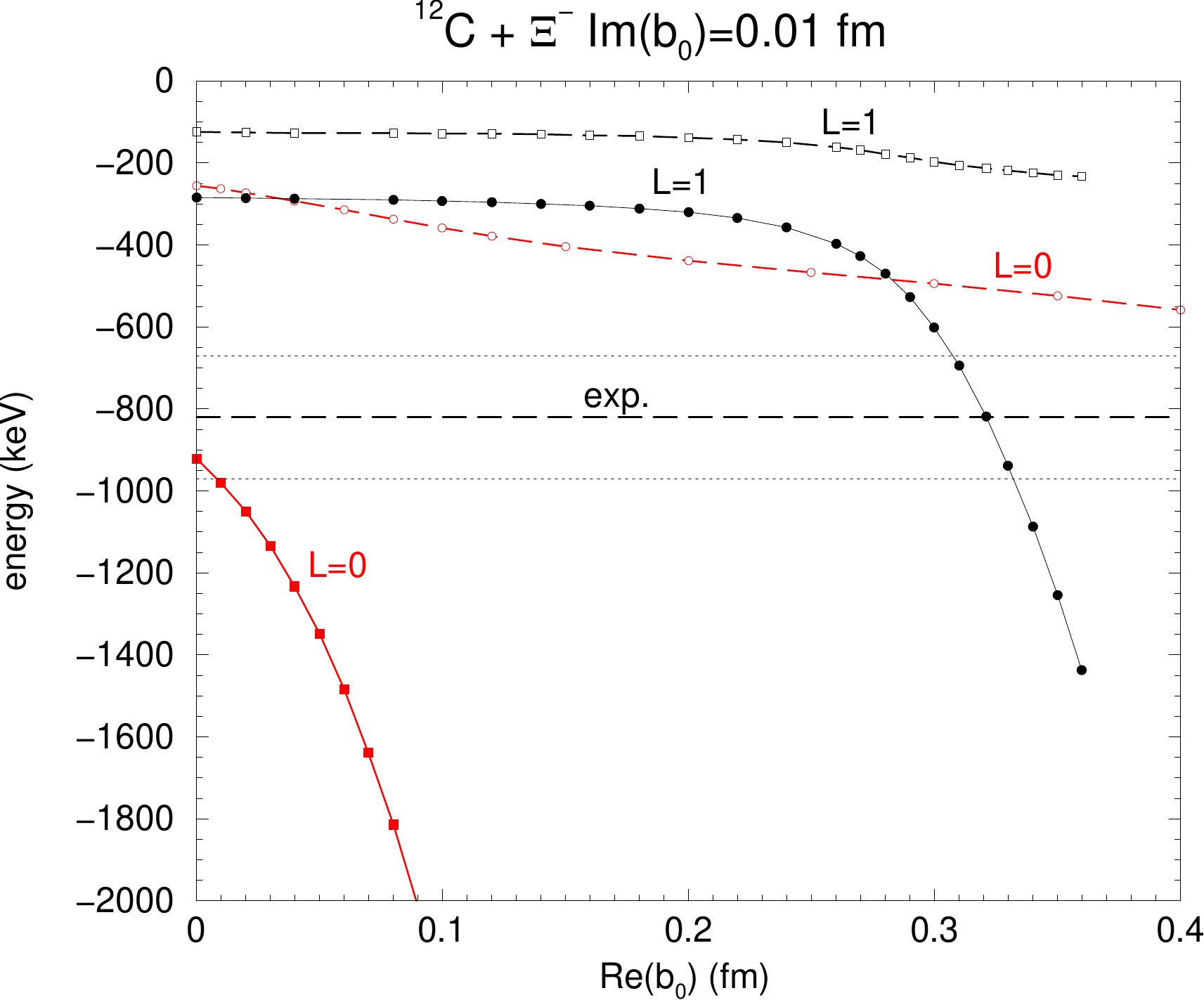} 
\caption{Energy levels (in keV) of the lowest $\Xi^-$ atomic states for 
L=0 (1S,2S) and L=1 (2P,3P) in $^{12}$C as a function of the strength 
parameter Re$\,b_0$ (in fm) of the $\Xi^-$ optical potential (\ref{eq:EEs}). 
Spin-orbit splittings of L=1 states are suppressed in this figure. The dashed 
and dotted horizontal lines mark the value $B_{\Xi^-}=0.82\pm 0.15$~MeV from 
Table~\ref{tab:events}.} 
\label{fig:Xi12C} 
\end{center} 
\end{figure}

\section{$\Xi^-$ capture in $^{12}$C} 
\label{sec:12C} 

Figure~\ref{fig:Xi12C} shows a portion of the combined atomic plus nuclear 
spectrum of $\Xi^-$ in $^{12}$C, $B_{\Xi^-}\leq 2$~MeV, as a function of 
Re$\,b_0$, Eq.~(\ref{eq:EEs}), for a fixed Im$\,b_0= 0.01$~fm corresponding 
to a nuclear-matter $\Xi^-$ capture width $\Gamma_{\Xi^-}\approx 1.5$~MeV, 
compatible with the HAL-QCD weak $\Xi^- p\to\Lambda\Lambda$ transition 
potential~\cite{HALQCD20}. Plotted are the energies of the two lowest states 
of each orbital angular momentum $l_{\Xi^-}=0,1$, starting at Re$\,b_0$=0 with 
almost purely atomic states 1S,2P,2S,3P from bottom up. Of these states the 
1S state with Bohr radius about 3.8~fm is indistinguishable from a nuclear 
$1s$ state, and indeed it dives down in energy as soon as Re$\,b_0$ is made 
nonzero. It takes considerable strength, Re$\,b_0\gtrsim 0.25$~fm, before 
the next atomic state, 2P with Bohr radius about 15~fm, overlaps appreciably 
with the $^{12}$C$_{\rm g.s.}$ nuclear core, diving down in energy to become 
a nuclear $1p$ state. The higher two states that start as atomic 2S,3P 
remain largely atomic as Re$\,b_0$ is varied in Fig.~\ref{fig:Xi12C}. 
Their slowly decreasing energies indicate a rearrangement of the atomic 
spectrum~\cite{GFB96}: 2S$\to$1S and 3P$\to$2P. Judging by the marked band 
of values $B_{\Xi^-}$=0.82$\pm$0.15~MeV for the two KEK E176 events listed in 
Table~\ref{tab:events}, the figure suggests that they are compatible with 
a $1p_{\Xi^-}$ nuclear state corresponding to a $\Xi$-nuclear potential 
strength of Re$\,b_0=0.32\pm 0.01$~fm. The sensitivity to variations of 
Im$\,b_0$ is minimal: choosing Im$\,b_0=0.04$~fm~\cite{BFG99} instead of 
0.01~fm increases Re$\,b_0$ by 0.01~fm to $0.33\pm 0.01$~fm. 

Radiative rates for E1 transitions from the $\Xi^-$ atomic 3D state to the 
$\Xi^-$ atomic 3P state, and to the $1p_{\Xi}$ nuclear state that started 
as atomic 2P state are found comparable to each other, accounting together 
for 7.6\% of the total 3D width $\Gamma_{\rm 3D}=3.93$~eV as obtained using 
the optical potential (\ref{eq:EEs}). However, the subsequent $\Xi^- p\to
\Lambda\Lambda$ capture will proceed preferentially from the $1p_{\Xi}$ 
nuclear state that offers good overlap between the $1p_{\Xi^-}$ and $1p_p$ 
valence-proton orbits. Since the final $1s_{\Lambda}^2$ configuration has 
$J_f=0$, and the $p$-shell protons in $^{12}$C are mostly in $j=\frac{3}{2}$ 
orbits, the requirement $J_i=J_f=0$ imposes $j_{\Xi^-}=\frac{3}{2}$ on the 
spin-orbit doublet members of the $1p_{\Xi^-}$ state. The shift of this 
$1p_{\Xi^-}(\frac{3}{2})$ sub level from the $1p_{\Xi^-}$ $(2j+1)$-average is 
estimated, based on the 152~keV $1p_{\Lambda}$ spin-orbit splitting observed 
in \lam{13}{C}~\cite{Ajimura01} to be less than 100~keV upward~\cite{MJ94} 
and, hence, within the 0.15~MeV listed uncertainty introduced here for the 
position of the $(2j+1)$-averaged $1p_{\Xi^-}$ state.

\section{Spectroscopy of $^{14}$N$_{\rm g.s.} + 1p_{\Xi^-}$ states} 
\label{sec:14N} 

Having derived the strength parameter Re$\,b_0=0.32\pm 0.01$~fm of 
$V_{\rm opt}$ by fitting it to $B_{\Xi^-}^{1p}(^{12}$C)=0.82$\pm$0.15~MeV, 
we apply this optical potential to $^{14}$N where it yields $B_{\Xi^-}^{1p}
(^{14}$N)=2.08$\pm$0.28~MeV, considerably higher than the value $B_{\Xi^-}
$=1.15$\pm$0.20~MeV obtained from the three events assigned in 
Table~\ref{tab:events} to $\Xi^-$ capture in $^{14}$N. However, this 
calculated $B_{\Xi^-}^{1p}(^{14}$N) corresponds to a $(2{\cal L}+1)$-average 
of binding energies for a triplet of states ${\cal L}^{\pi}=(0^-,1^-,2^-)$ 
obtained by coupling a $1p_{\Xi^-}$ state to $J^{\pi}(^{14}$N$_{\rm g.s.})
$=1$^+$, as shown in Fig.~2. We now discuss the splitting of these triplet 
states. Effects of $\Xi^-$ Pauli spin, $s_{\Xi^-}=\frac{1}{2}$, are introduced 
at a later stage. 

\begin{figure}[htb] 
\begin{center} 
\includegraphics[width=0.8\textwidth]{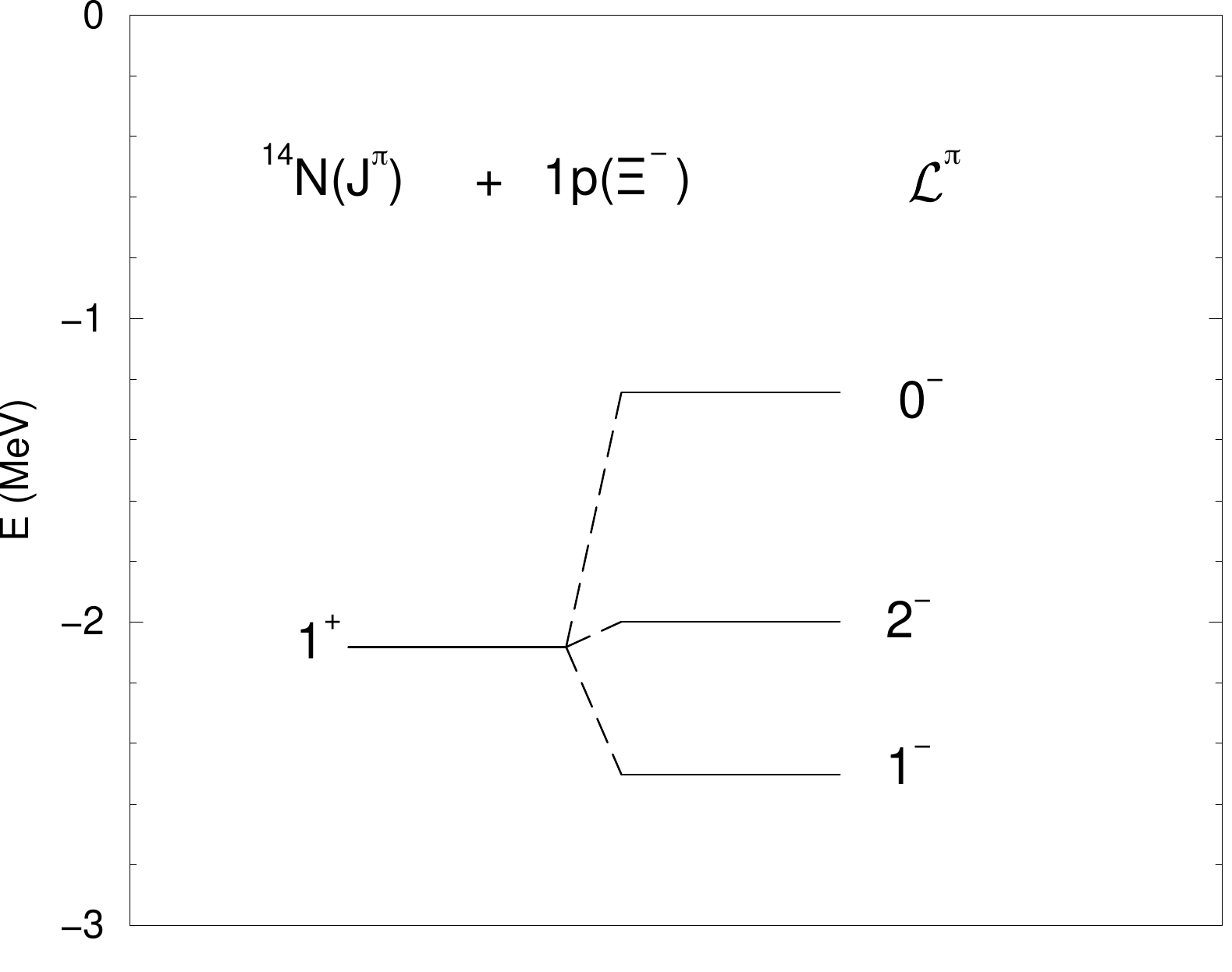} 
\caption{Energies (in MeV) of ${\cal L}^{\pi}=(0^-,1^-,2^-)$ triplet of 
$^{14}$N$_{\rm g.s.} + 1p_{\Xi^-}$ states, split by a $Q_N\cdot Q_{\Xi}$ 
residual interaction (\ref{eq:QdotQ}). The $(2{\cal L}+1)$-averaged energy 
$-2.08\pm 0.28$~MeV was calculated using the same optical potential 
parameter $b_0$ that yields a $^{12}$C$_{\rm g.s.} + 1p_{\Xi^-}$ state at 
$-0.82\pm 0.15$~MeV, corresponding to the $\Xi^-$ capture events in $^{12}$C 
listed in Table~\ref{tab:events}.}   
\label{fig:Xi14N} 
\end{center} 
\end{figure} 

The construction of the $^{14}$N$_{\rm g.s.} + 1p_{\Xi^-}$ spectrum in 
Fig.~\ref{fig:Xi14N} follows a similar $^{12}$C(2$^+$;4.44~MeV)$\,+ 
1p_{\Lambda}$ spectrum construction in \lam{13}{C}~\cite{Auerbach81}. 
The energy splittings marked in the figure are obtained from a two-body 
spin-independent interaction $V_0(r_{\Xi N})$, Eq.~(\ref{eq:V2body}), 
between a $p$-shell $\Xi$ hyperon and $p$-shell nucleons, expressed in 
terms of its shell-model (SM) quadrupole-quadrupole residual interaction 
${\cal V}_{\Xi N}$, 
\begin{equation} 
{\cal V}_{\Xi N}=F^{(2)}_{\Xi N} Q_N\cdot Q_{\Xi}, \,\,\,\,\,\, 
Q_B=\sqrt{\frac{4\pi}{5}} Y_2({\hat{r}}_B), 
\label{eq:QdotQ}
\end{equation} 
where $F^{(2)}$ is the corresponding Slater integral~\cite{dST63}. 
A representative value of $F^{(2)}_{\Xi N}=-3$~MeV is used here, smaller than 
the value $F^{(2)}_{\Lambda N}=-3.7$~MeV established empirically for $p$-shell 
$\Lambda$ hypernuclei~\cite{DalGal81}, in accordance with a $\Xi N$ strong 
interaction somewhat weaker than the $\Lambda N$ strong interaction 
(see next section). A single $^3D_1$ $^{14}$N$_{\rm g.s.}$ SM component 
providing a good approximation to the full SM intermediate-coupling g.s. 
wavefunction~\cite{Mil07} was assumed in the present evaluation. 

Fig.~\ref{fig:Xi14N} exhibits a triplet of $^{14}$N$_{\rm g.s.} + 1p_{\Xi^-}$ 
levels, spread over more than 1~MeV. The least bound triplet state, with 
${\cal L}^{\pi}=0^-$, is shifted upward by 0.84~MeV from the $(2{\cal L}+1)$ 
averaged position at $-2.08\pm 0.28$~MeV to $E(0^-)=-1.24\pm 0.28$~MeV. This is 
consistent with the averaged position $\bar{E}=-1.15\pm 0.20$~MeV of the three 
$\Xi^-\,^{14}$N$_{\rm g.s.}$ capture events listed in Table~\ref{tab:events}. 
We are not aware of any good reason why capture has not been seen from the 
other two states with ${\cal L}^{\pi}=1^-,2^-$. This may change when more 
events are collected at the next stage of the ongoing J-PARC E07 emulsion 
experiment. 

\begin{table}[htb] 
\begin{center} 
\caption{Quadrupole-quadrupole contributions to the energies 
$E({\cal L}^{\pi})$ of the $^{14}$N$_{\rm g.s.}+1p_{\Xi^-}$ triplet of 
states shown in Fig.~\ref{fig:Xi14N} with respect to $E(^{14}$N$_{\rm g.s.})$, 
using $F^{(2)}_{\Xi N}=-3$~MeV, and spin contributions to the splittings 
$\Delta E({\cal L})=E(J={\cal L}+\frac{1}{2})-E(J={\cal L}-\frac{1}{2})$ 
of the ${\cal L}\neq 0$ states. $A_{ls}$ and $A_{ss}$ are spin-orbit 
($l_{\Xi}=1,s_{\Xi}=\frac{1}{2}$) and spin-spin ($s_N=s_{\Xi}=\frac{1}{2}$) 
energy splittings, respectively, see text.} 
\begin{tabular}{cccccc}
\hline
Interaction & $E(0^-)$ & $E(1^-)$ & $E(2^-)$ & $\Delta E(1^-)$ &
$\Delta E(2^-)$  \\
\hline
$Q_N\cdot Q_{\Xi}$ & $-\frac{7}{25}\,F^{(2)}_{\Xi N}$ &
$\frac{7}{50}\,F^{(2)}_{\Xi N}$ & $-\frac{7}{250}\,F^{(2)}_{\Xi N}$ &
-- & --  \\
$l_{\Xi}\cdot s_{\Xi}$ & -- & -- & -- & $\frac{1}{2}\,A_{ls}$ &
$\frac{5}{6}\,A_{ls}$  \\
$s_N\cdot s_{\Xi}$     & -- & -- & -- & $\frac{3}{8}\,A_{ss}$ &
$\frac{5}{8}\,A_{ss}$  \\
\hline
\end{tabular}
\label{tab:spin}
\end{center}
\end{table}

Introducing Pauli spin, $s_{\Xi^-}=\frac{1}{2}$, the total angular momentum of 
the uppermost level marked $0^-$ in Fig.~\ref{fig:Xi14N} becomes $J^{\pi}=
{\frac{1}{2}}^-$, but its position is unaffected by spin-orbit and spin-spin 
interactions. Each of the other two levels in Fig.~\ref{fig:Xi14N} splits into 
a doublet $J={\cal L}\pm\frac{1}{2}$ whose $(2J+1)$-average remains in the 
unsplit position. Estimated splittings are listed in Table~\ref{tab:spin} 
in terms of two constituent spin matrix elements: $A_{ls}\lesssim 
300$~keV~\cite{MJ94} for the $l_{\Xi}\cdot s_{\Xi}$ spin-orbit splitting 
$E_{1p_{\Xi}}({\frac{3}{2}}^-)-E_{1p_{\Xi}}({\frac{1}{2}}^-)$, and $A_{ss}
\approx 400\pm 80$~keV for the $s_N\cdot s_{\Xi}$ spin-spin splitting 
$E(S_{\Xi N}=0)-E(S_{\Xi N}=1)$ for $p$-shell nucleon and $\Xi$ hyperon. 
For estimating $A_{ss}$ we used the HAL-QCD~\cite{HALQCD20} volume integral 
of $V_{\sigma}$, Eq.~(\ref{eq:V2body}), relative to that of $V_0$, 
thereby generating about 20\% systematic uncertainty. Incorporating these 
spin splittings into the ${\cal L}^{\pi}=(0^-,1^-,2^-)$ triplet in 
Fig.~\ref{fig:Xi14N} keeps the ${\cal L}^{\pi}=0^-$ state of interest, which 
has become $J^{\pi}={\frac{1}{2}}^-$, well separated by at least 0.5~MeV from 
the rest of the split states.

\section{Density dependence, $\Xi$ nuclear potential depth and $1s_{\Xi^-}$ 
states} 
\label{sec:light} 

So far we have discussed a density independent $t$-matrix element $b_0$ 
in $V_{\rm opt}$, Eq.~(\ref{eq:EEs}), to fit the $\Xi^-$ capture 
events in $^{12}$C from Table~\ref{tab:events}. The resulting value 
Re$\,b_0=0.32\pm 0.01$~fm implies, in the limit $A\to\infty$ and 
$\rho(r)\to\rho_0$=0.17~fm$^{-3}$, a value $V_{\Xi}=24.3\pm 0.8$~MeV 
in nuclear matter, in accordance with the extraction of $V_{\Xi}$ from 
old emulsion events~\cite{DG83} but exceeding considerably other values 
reviewed in the Introduction. To explore how robust this conclusion is, 
we introduce the next to leading-order density dependence of $V_{\rm opt}$, 
replacing Re$\,b_k$ ($k$=0,1) in Eq.~(\ref{eq:EEs}) by 
\begin{equation} 
{\rm Re}\,b_k(\rho)=\frac{{\rm Re}\,b_k}
{1+\frac{3k_F}{2\pi}{\rm Re}\,b_0^{\rm lab}},\,\,\,\,\,\,\,\,\,\,
k_F=(3{\pi}^2\rho/2)^{\frac{1}{3}}, 
\label{eq:WRW} 
\end{equation} 
where $k_F$ is the Fermi momentum corresponding to nuclear density $\rho$ 
and $b_0^{\rm lab}=(1+\frac{m_{\Xi^-}}{m_N})b_0$ is the lab transformed 
form of the c.m. scattering amplitude $b_0$. Eq.~(\ref{eq:WRW}) 
accounts for Pauli exclusion correlations in $\Xi N$ in-medium multiple 
scatterings~\cite{DHL71,WRW97}. Variants of the form (\ref{eq:WRW}) have been 
used in kaonic atoms~\cite{FG13} and mesic nuclei~\cite{WH08,FGM13,CFGM14} 
calculations. Shorter-range correlations, disregarded here, were shown in 
Ref.~\cite{WH08} to contribute less than $\sim$30\% of the long-range Pauli 
correlation term. Applying Eq.~(\ref{eq:WRW}) in the present context, 
$B_{\Xi^-}^{1p}(^{12}$C)=0.82~MeV is refitted by Re$\,b_0$=0.527~fm, 
lowering the former value $B_{\Xi^-}^{1p}(^{14}$N)=2.08~MeV to 1.95~MeV 
without any substantive change in the conclusions drawn above regarding 
the five two-body capture events deciphered here. The nuclear-matter 
$\Xi$-nuclear potential depth $V_{\Xi}$ decreases from 24.3$\pm$0.8 to 21.9$
\pm$0.7~MeV, a decrease of merely 10\%, with additional systematic uncertainty 
of less than 1~MeV. This value of $V_{\Xi}$ is sufficient to bind $1s_{\Xi^-}$ 
states in $p$-shell nuclei, with systematic uncertainty of less than 0.5~MeV, 
as demonstrated in Table~\ref{tab:1s} which shows a steady decrease of 
$B_{\Xi^-}^{1s}$ and $\Gamma_{\Xi^-}^{1s}$ down to $^4$He. The increased 
$\Gamma_{\Xi^-}^{1s}(^4$He) reflects a denser $^4$He medium. However, 
expecting corrections of order ${\cal O}(1/A)$ to the optical potential 
methodology, our $^4$He result should be taken with a grain of salt. 
It is worth noting that all listed $1s_{\Xi^-}$ g.s. levels remain bound 
also when the attractive finite-size Coulomb interaction $V_c$ is switched 
off. None of such $1s_{\Xi^-}$ states have been observed conclusively in 
dedicated experiments. 

\begin{table}[ht!]
\begin{center}
\caption{Binding energies $B_{\Xi^-}^{1s}$ and widths $\Gamma_{\Xi^-}^{1s}$
(in MeV) in core nuclei $^{A}$Z($J_c$), g.s. spin $J_c$, obtained by solving
Eq.~(\ref{eq:KG}) with $b_0=0.527+{\rm i}\,0.010$~fm and $b_1=-0.225$~fm in
$V_{\rm opt}$, Eqs.~(\ref{eq:EEs}),(\ref{eq:WRW}). A finite-size Coulomb
interaction $V_c$ is included.} 
\begin{tabular}{crrrrrr}
\hline 
 & $^{14}$N(1) & $^{12}$C(0) & $^{11}$B($\frac{3}{2}$) & $^{10}$B(3) &
$^{6}$Li(1) & $^{4}$He(0)  \\
\hline 
$B_{\Xi^-}^{1s}$ & 11.5 & 9.8 & 8.4 & 7.6 & 2.1 & 2.0  \\
$\Gamma_{\Xi^-}^{1s}$ & 1.02 & 0.93 & 0.89 & 0.77 & 0.26 & 0.45  \\
\hline 
\end{tabular} 
\label{tab:1s} 
\end{center} 
\end{table} 

The $T=\frac{1}{2}$ $^{11}$B nucleus, the only $T\neq 0$ nucleus listed 
in Table~\ref{tab:1s}, requires in addition to the isoscalar parameter 
$b_0=0.527+{\rm i}\,0.010$~fm also a knowledge of the isovector parameter 
$b_1$. Here we used the HAL-QCD~\cite{HALQCD20} volume integral 
of $V_{\tau} (r_{\Xi N})$ relative to that of $V_0(r_{\Xi N})$, 
Eq.~(\ref{eq:V2body}), to estimate $b_1$ relative to $b_0$, thereby deriving 
a value $b_1=-0.225$~fm.{\footnote{Note that Im$\,b_1$=0 because the charge 
exchange $\Xi^-$+$^{11}$B$\to$$\Xi^0$+$^{11}$Be is kinematically blocked.}} 
The resulting value $B_{\Xi^-}^{1s}(^{11}$B) listed in the table is lower by 
530 keV than obtained disregarding $b_1$. Next, we introduce $s_{\Xi^-}$=$ 
\frac{1}{2}$ Pauli spin, splitting each of the listed $1s_{\Xi^-}$ levels 
in $J_c\neq 0$ core nuclei into two sub levels $J=J_c\pm\frac{1}{2}$. Using 
HAL-QCD~\cite{HALQCD20} ratios of volume integrals of $V_{\sigma}(r_{\Xi N})$ 
and $V_{\sigma\tau}(r_{\Xi N})$ to that of $V_0(r_{\Xi N})$, 
Eq.~(\ref{eq:V2body}), as done above for $V_{\tau}$, we estimate the 
$1s_{\Xi^-}$ spin splittings to be well below 1~MeV. Other potential 
sources of $\Xi^-$ spin splittings that are relevant in $\Lambda$ hypernuclei, 
such as tensor or induced nuclear spin-orbit terms, are likely to be 
considerably weaker than evaluated in Ref.~\cite{Mil12} and are disregarded 
here. Of particular interest is the ($^{11}$B$_{\rm g.s.}+1s_{\Xi^-}$) 
$J^{\pi}=1^-$ doublet member expected to be formed in the $^{12}$C($K^-,K^+$) 
production reaction when the outgoing $K^+$ meson is detected in the forward 
direction. Our estimates place it about 0.5~MeV deeper than the listed 
$(2J+1)$-averaged $B_{\Xi^-}^{1s}(^{11}$B)=8.4~MeV, contrasting statements, 
e.g.~\cite{HN18}, that adopt $B_{\Xi^-}^{1s}(^{11}$B)$\sim$5~MeV from the BNL 
AGS-E885 $^{12}$C($K^-,K^+$) experiment~\cite{E885}. In fact, the E885 poor 
resolution prevents making any such conclusive statement.

\section{Conclusion} 
\label{sec:concl} 

We have shown that all five light nuclear emulsion events identified in KEK 
and J-PARC $K^-$ exposure experiments as two-body $\Xi^-$ capture in $^{12}$C 
and $^{14}$N into twin $\Lambda$ hypernuclei correspond to capture from 
$1p_{\Xi^-}$ Coulomb-assisted bound states. This involved using just one 
{\it common} strength parameter of a density dependent optical potential. 
Long-range $\Xi N$ shell-model correlations were essential in making the 
$^{14}$N events consistent with the $^{12}$C events. Earlier attempts to 
explain these data overlooked this point, therefore reaching quite different 
conclusions~\cite{yam01,hiyama16,sun16,hu17,jin20}. Predicted then are 
$1s_{\Xi^-}$ bound states with $B^{1s}_{\Xi^-}\sim 10$~MeV in $^{12}$C and 
somewhat larger in $^{14}$N, deeper by 4--5 MeV than the $1s_{\Xi^-}$ states 
claimed by a recent J-PARC E07 report of multibody capture events~\cite{E07b}. 
The $\Xi$ nuclear-matter potential depth derived here within an optical 
potential methodology, $V_{\Xi}=21.9\pm 0.7$~MeV, is considerably larger 
than $G$-matrix values below 10~MeV derived from recent LQCD and EFT 
$\Xi N$ potentials~\cite{HALQCD19,K19}. A systematic optical-potential 
model uncertainty of less than 1~MeV as discussed in Sect.~\ref{sec:light} 
is short of bridging the gap noted above. Substantial $\Xi NN$ three-body 
{\it attractive} contributions to the $\Xi$ nuclear potential depth would 
be required to bridge this gap. Intuitively one expects repulsive $BNN$ 
three-body contributions for octet baryons $B$, e.g. Ref.~\cite{ESC16}, 
but in chiral EFT studies, focusing on decuplet-$B^{\ast}$ intermediate 
$B^{\ast}NN$ and $BN\Delta$ configurations, this has been proven so far 
only for $B=\Lambda$~\cite{PHKMW17,GKW20}. 

To check the procedure practised in Sect.~\ref{sec:light} for 
fitting $V_{\Xi}$ to just one $\Xi^-$-$^{12}$C bound state datum, 
we apply it to the $1s_{\Lambda}$ binding energy in $^{12}$C, 
$B_{\Lambda}^{1s}=11.69\pm 0.12$~MeV \cite{Davis05}. The fitted strength 
$b_0=0.866\pm 0.010$~fm amounts to a nuclear-matter $\Lambda$ potential depth 
$V_{\Lambda}=31.7\pm 0.2$ MeV, in good agreement with the accepted value 
$V_{\Lambda}\approx 30$ MeV \cite{GHM16}. One may slightly improve the derived 
value of $V_{\Lambda}$ by subtracting from $B_{\Lambda}^{1s}$ a nuclear 
induced spin-orbit contribution that vanishes in the limit $A\to\infty$, 
thereby reducing our input $B_{\Lambda}^{1s}$ to 10.85~MeV~\cite{Mil12}. 
This gives $b_0=0.798\pm 0.010$~fm and $V_{\Lambda}=30.5\pm 0.2$ MeV. 
Here too it is not possible to separate the contribution of $\Lambda NN$ 
three-body potential terms from that of the main $\Lambda N$ two-body 
potential term. 

A strong $\Xi$-nuclear interaction, such as derived here, may have 
far-reaching implications to ($N,\Lambda,\Xi$) strange hadronic 
matter~\cite{Schaffner00} and particularly to dense neutron star 
matter~\cite{Schaffner12}. In the latter case a strong $\Xi$-nuclear 
interaction might cause a faster depletion of $\Lambda$ hyperons by 
$\Lambda\Lambda\to \Xi^- p$, a process inverse to the $\Xi^-$ capture 
reaction considered in the present work. More work is necessary in this 
direction.

\section*{Acknowledgments} 
Recent related correspondence with Johan Haidenbauer, Ji\v{r}\'{i} Mare\v{s}, 
John Millener, Tomofumi Nagae, Josef Pochodzalla and Tom Rijken is gratefully 
acknowledged. The present work is part of a project funded by the European 
Union's Horizon 2020 research \& innovation programme, grant agreement 824093.

\end{document}